\documentclass{article}       
\usepackage{e}                
\usepackage[dvips]{graphicx}
\begin{document}
\newcommand{\ox}{ $\rightarrow$ }
\newcommand{\qqb}{$q \overline{q}$ }
\newcommand{\emiss}{$\not\hskip-5truedd E_{T}$ }
\newcommand{\ptmiss}{$\not\hskip-5truedd P_{T}$ }
\newcommand{\pT}{$p_{T}$ }
\newcommand{\X}{\times}
\newcommand{\lov}{\overline{l}}
\branch{C}   
%
\title{Production and Decay of Excited Electrons at the LHC}
\author{O.\c{C}ak\i r\inst{1}\and C. Leroy\inst{2}\and R. Mehdiyev\inst{2,3}\and 
        A. Belyaev$^{4,5}$}
\institute{ 
Ankara University, Faculty of Sciences, Department of Physics, 06100, Tandogan, Ankara, Turkey. \and
Universit\'e de Montr\'eal, D\'epartement de Physique, Montr\'eal, H3C 3J7, Canada.\and
on leave of absence from Institute of Physics, Azerbaijan National Academy of Sciences, 
370143, Baku, Azerbaijan.\and
 Department of Physics, Florida State University, Tallahassee, FL, USA.\and
on leave of absence from Skobeltsyn Institute for Nuclear Physics,
Moscow State University, 119 899, Moscow, Russian Federation}
\maketitle
   \begin{abstract}
We study single production of excited electrons at the CERN LHC through 
contact interactions of fermions. 
Subsequent decays of excited electrons to ordinary electrons
and light fermions via gauge and contact interactions are examined.
The mass range accessible with the ATLAS detector is obtained. 
\end{abstract}
%
\section{Introduction}
One of the  natural explanations for the existence of the fermionic 
generations is the compositeness\cite{1} of the known leptons and quarks, 
sharing common constituents (preons). According to this approach, 
a quark or lepton might be a bound state of three fermions \cite{2} 
or a fermion and a boson \cite{3}. 
In many models along this line, quarks and leptons are composed
of a scalar and a spin-1/2 preon. 
Composite models \cite{1} predict a rich spectrum of excited 
states\cite{1,4} of known particles.
Naturally, in these models spin-1/2 is assigned to the lowest lying radial 
and orbital excitations. Searching for the spin of the lowest lying 
fermionic excitations could thus give a rather direct clue to the underlying
 preon structure. Therefore, it is important to conduct experiments which 
will probe possible substructure of leptons and quarks and test the variety 
of preonic models.

Exchange of preons may lead to contact interactions between quarks
and leptons. In this sense, it is conceivable that the standard model (SM) is
just the low energy limit of a more fundamental theory which is characterized
by a large mass scale $ \Lambda $. 
The existence of four-fermion contact interactions would be a signal of 
new physics beyond the SM.
If the experimental energy scale is high enough, the nature of this new 
physics can be probed. It is expected that the next generation of hadron 
colliders like the LHC which will achieve very high centre of mass energies 
will extend the search for composite states. 
In particular, contact interactions may be an important source for excited 
lepton production at the CERN LHC. 
Updated lower limits on the scale of compositeness $ \Lambda $ are given by 
the ALEPH Collaboration, $ \Lambda > 6.2$ TeV \cite{5} and 
D0 Collaboration $ \Lambda > 4.2$ TeV \cite{6}. Lower mass limits for 
excited lepton mass are given by 
the ZEUS Collaboration, $m_{e^{\star}} > 200$ GeV\cite{7} and 
the OPAL Collaboration, $m_{e^{\star}} > 306$ GeV\cite{8}.   

In the present study, single production of excited electrons is simulated through contact interactions at LHC energies.

The excited muon production is similar to $e^{\star}$ production at hadron colliders. However, with a good $e/\mu$ separation, $\mu^{\star}$ can be easily detected. But overall statistical significances for 
$\mu^{\star}$ should be very similar to an excited electron production case.

This work is a continuation of previous work devoted to the study of 
the excited quark production with subsequent decay to a quark and photon 
\cite{9}, a quark and gluon\cite{10} and a quark and $ W/Z $\cite{11}.  

\section{Effective Lagrangian}
In phenomenological models, it is assumed that any theory of compositeness at
large mass scale must have a low energy limit that preserves the symmetries
of the SM. 
If quarks and leptons are composite at the energy scale $\Lambda$, the strong
forces binding their constituents induce flavor-diagonal contact
interactions, which have significant effects at subprocess energies well
below $\Lambda$\cite{4}. Contact interactions between quarks and leptons may
appear as the low energy limit of the exchange of heavy particles.
At sufficiently high energies excited fermions could be produced directly.
They should form weak iso-doublets and carry
electromagnetic charges similar to those of the ordinary fermions. 
We will assume that the excited leptons have spin and isospin 1/2 to limit 
the number of parameters.

\subsection{Production of excited electrons}
Excited electrons may couple to ordinary quarks via contact interactions 
resulting from preon interactions. 
For energies below the compositeness scale $\Lambda$, these interactions
can be described by an effective four-fermion Lagrangian\cite{4}

\begin{equation}
\label{1}
L_{C}=\frac{g_{\star }^{2}}{2\Lambda ^{2}}j^{\mu }j_{\mu }
\end{equation}
 with 

\begin{equation}
\label{2}
j_{\mu }=\eta _{L}\overline{f}_{L}\gamma _{\mu }f_{L}+\eta _{L}^{\prime }\overline{f}_{L}^{\star }\gamma _{\mu }f_{L}^{\star }+\eta _{L}^{\prime \prime }\overline{f}_{L}^{\star }\gamma _{\mu }f_{L}+(L\rightarrow R)+h.c.
\end{equation}
 where the coupling $g_{\star }^{2}=4\pi$; $\eta _{L}$ and $\eta _{R}$
are coefficients for left-handed and right-handed currents, respectively. 
Here we assume $\eta_{L}=\eta_{L}^{\prime}=\eta _{L}^{\prime \prime }=1$ and 
$\eta _{R}=\eta_{R}^{\prime}=\eta _{R}^{\prime \prime }=0$, for simplicity.

\begin{figure}
{\par\centering 
\resizebox*{0.6\textwidth}{0.17\textheight}{\includegraphics{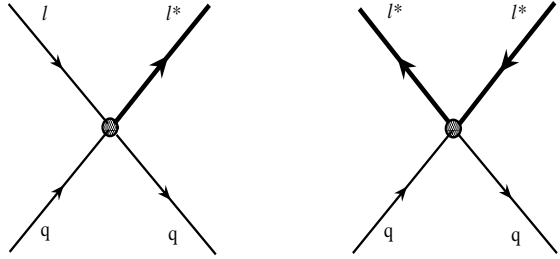}} \par}
\caption{Contact interactions of excited electrons ($l^{\star}$) with quark ($q$) and 
leptons ($l$)}
\label{fig1}
\end{figure}

At the CERN LHC pp collider, excited electrons can be produced either singly 
$q\overline{q}\rightarrow l\lov^{\star},l^{\star }\overline{l}$
or in pairs $q\overline{q}\rightarrow l^{\star }\overline{l}^{\star }$
through contact interactions as shown in Fig.~\ref{fig1}. 

The partonic cross sections for excited electron production through contact interactions are given by \cite{4}

\begin{equation}
\label{10}
\widehat{\sigma }(q\overline{q}\rightarrow l\overline{l}^{\star },l^{\star }\overline{l})=\frac{\pi }{6\widehat{s}}\left( \frac{\widehat{s}}{\Lambda ^{2}}\right) ^{2}\left( 1+\frac{v}{3}\right) \left( 1-\frac{m_{\star}^{2}}{\widehat{s}}\right) ^{2}\left( 1+\frac{m_{\star}^{2}}{\widehat{s}}\right) 
\end{equation}

\begin{equation}
\label{11}
\widehat{\sigma }(q\overline{q}\rightarrow l^{\star }\overline{l}^{\star })=\frac{\pi \widetilde{v}}{12\widehat{s}}\left( \frac{\widehat{s}}{\Lambda ^{2}}\right) ^{2}\left( 1+\frac{\widetilde{v}^{2}}{3}\right) 
\end{equation}
 where 

\begin{equation}
\label{12}
v=\frac{\widehat{s}-m_{\star}^{2}}{\widehat{s}+m_{\star}^{2}}\, \, ,\quad \widetilde{v}=\left( 1-4\frac{m_{\star}^{2}}{\widehat{s}}\right) ^{1/2}
\end{equation}
where $\widehat{s}$ denotes the Mandelstam variable for the subprocess, the centre of mass energy.

\begin{figure}
{\par\centering 
\resizebox*{0.8\textwidth}{0.4\textheight}{\includegraphics{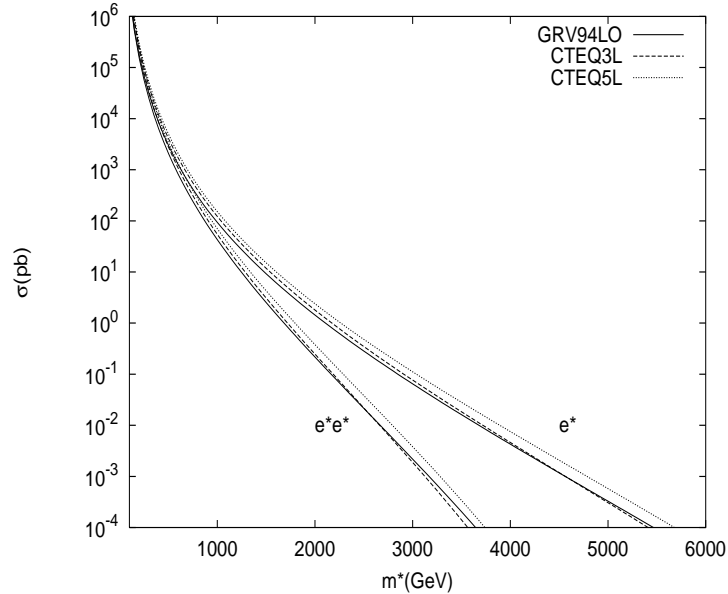}} \par}
\caption{Cross sections for single and pair production of excited electrons through contact interactions at LHC ($\Lambda=m_{\star}$) calculated using different PDFs}
\label{fig2}
\end{figure}

Cross sections for single and pair production of excited electrons through 
contact interactions are shown in Fig.~\ref{fig2}.
Since $ l^{\star }\overline{l^{\star }} $ pair production requires larger
centre of mass energy than single $ l\overline{l}^{\star } $ production,
it is less favored.

Cross section values presented in this paper were calculated using CTEQ5L parton distribution 
function (PDF)\cite{12}. Other PDF's were tried: GRV94 LO\cite{13} and CTEQ3L\cite{14} 
differ in their cross-section prediction by about 20\% and 30\%, respectively. 
These PDF's extend slightly the mass reach for excited electrons at the LHC.

One should notice that excited leptons can also be  produced via gauge  interactions (see
the effective Lagrangian in the next subsection).  Gauge interactions can give rise to
$\ell^*\ell^*$, $\ell\ell^*$, $\ell^*\nu$ signatures.
The study of such scenario has been done in ~\cite{15}. However,
since those processes involve electromagnetic or electroweak couplings they
contribute to less than 1\%  compared to the excited lepton production rate
via contact interactions.

In the same time contact interactions will modify the Drell-Yan(DY) dilepton production 
process and one should definitively check how big are deviations from
the SM~\cite{15}.
It is important to notice that the sensitivity of the LHC to the compositeness scale in 
the Drell-Yan channel is very high -- for an integrated luminosity
of 100 $fb^{-1}$,  one has $\Lambda<30$~TeV at 5 sigma level limit !~\cite{22}

This means that, for the range of $\Lambda \sim 30$~TeV we consider here,
the  signal in the DY channel from contact interactions will be unavoidably large !  
Therefore our study should be considered as the study not for the determination of 
the limits on $\Lambda$
(which can be done {\it much} better using DY channel) but as the study of the excited
electron production at the LHC. One should stress the big complementary
role of the DY channel, however, if one would observe signal from
the excited electron production, the signal from contact interactions
should also be present in DY channel too.

One should notice also that the "charged current" contact term like 
$u d$ \ox $e \nu$ is not forbidden by $U(1) \times SU(2)L \times SU(3)$ 
symmetry and can give  rise to $l^{*}\nu$ production in contact interactions 
at the LHC. We do not consider this effective interaction in our study.

\subsection{Decay via gauge interactions}
The effective Lagrangian which describes the transition between
excited and ground states via gauge interactions is given by \cite{4}

\begin{equation}
L_{G}=\frac{1}{2\Lambda }\overline{l}_{R}^{\star }\sigma ^{\mu \nu
}(gf\frac{ \tau }{2}W_{\mu \nu }+g^{\prime }f^{\prime
}\frac{Y}{2}B_{\mu \nu })l_{L}+h.c.
\end{equation}
where $W_{\mu \nu }$ and $B_{\mu \nu }$ are the field strength
tensors of the SU(2) and U(1) gauge fields with the gauge
structure constants $\tau $ and $Y$, respectively; the factors $f$ and
$f^{\prime}$ describe the effective deviations from the SM coupling constants;
$g$ and $g^{\prime}$ are the corresponding gauge coupling constants.
An excited electron can then decay via gauge interactions to a gauge boson and 
an ordinary lepton.

\begin{figure}
{\par\centering \resizebox*{0.3\textwidth}{0.15\textheight}{\includegraphics{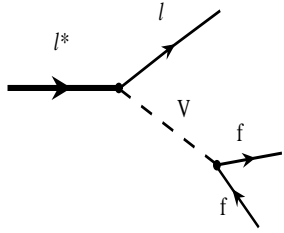}} \par}
\caption{Decay of excited lepton ($l^{\star }$) into a lepton ($l$) and a pair of
fermions ($f$) via gauge interactions mediated by the vector boson (V).}
\label{fig3}
\end{figure}

Here we assume that excited electrons have a mass larger than the $W$ and $Z$ boson 
masses and the main decay mode via gauge interaction will be two-body decays 
(Fig.~\ref{fig3}).

The decay widths:

\begin{eqnarray}
\Gamma _{G}(e^{\ast } &\rightarrow &\gamma e)=\frac{\alpha
_{\gamma }}{4}
\frac{m_{\star }^{3}}{\Lambda ^{2}}f_{\gamma }^{2} \\
\Gamma _{G}(e^{\ast } &\rightarrow &W\nu )=\frac{\alpha
_{W}}{4}\frac{ m_{\star}^{3}}{\Lambda
^{2}}f_{W}^{2}(1-\frac{m_{W}^{2}}{m_{\star }^{2}}
)^{2}(1+\frac{m_{W}^{2}}{2m_{\star }^{2}}) \\
\Gamma _{G}(e^{\ast } &\rightarrow &Ze)=\frac{\alpha
_{Z}}{4}\frac{m_{\star}^{3}}{\Lambda
^{2}}f_{Z}^{2}(1-\frac{m_{Z}^{2}}{m_{\star }^{2}})^{2}(1+\frac{
m_{Z}^{2}}{2m_{\star }^{2}})
\end{eqnarray}
where $f_{\gamma }=fT_{3}+f^{\prime }Y/2,$ $f_{W}=f/\sqrt{2}$ and $%
f_{Z}=fT_{3}\cos ^{2}\theta _{W}-f^{\prime }(Y/2)\sin ^{2}\theta
_{W},$ and $ T_{3}$ and $Y$ denotes the third component of the
weak isospin and hypercharge of $e^{\star },$ respectively. The
structure constants are defined as $\alpha_W=\alpha/\sin\theta_{\it{w}}$
and $\alpha_Z=\alpha_W/\cos\theta_{\it{w}}$, where $\theta_{\it{w}}$ is
the Weinberg angle.  For $m_{\star}\gg m_W,m_Z$ we
neglect $m_{W}^{2}/m_{\star }^{2}$ and $m_{Z}^{2}/m_{\star}^{2}$ terms, 
and the total width for gauge interaction decay can be obtained as follows

\begin{equation}
\Gamma _{G}(e^{\ast }\rightarrow
all)\simeq\frac{1}{4}\frac{m_{\star }^{3} }{\Lambda ^{2}}\left(
\alpha _{\gamma }f_{\gamma }^{2}+\alpha _{W}f_{W}^{2}+\alpha
_{Z}f_{Z}^{2}\right)
\end{equation}
where the parameters $f_{\gamma},f_{W}$ and $f_{Z}$ simply reduce
to the approximate values -1(0), 0.707(0.707) and -0.269(-0.5) for
$f=f^{\prime}=1$ ($f=-f^{\prime}=-1$), respectively.

The present analysis will assume $f=f^{\prime}=1$, but the results
can easily be reinterpreted for different values of these parameters, 
accounting for the change in branching ratio and intrinsic width.

\subsection{Decay via contact interactions}
The decay width of a fermion via contact interactions ( three-body decay ) is given by

\begin{equation}
\Gamma _{C}(e^{\star }\rightarrow e+f\overline{f})=\frac{1}{96\pi }N_{C}\ S\
\frac{m_{\star }^{5}}{\Lambda ^{4}}
\end{equation}

where $N_{C}$ is the number of colours of the fermions ($N_{C}=3(1)$ for
quarks(leptons)) and $S$ is a factor given by

\begin{eqnarray*}
S = 1 &for& f = q, l \neq e \\
S = 2 &for& f = e \\
\end{eqnarray*}

We present decay widths and branching ratios for gauge and contact
interactions in Table~\ref{table1},
for $\Lambda=m_{\star}$ and 
for the case when we fix $\Lambda$ at 6~TeV scale.
The  choice of  $\Lambda= 6$~TeV is motivated by the  fact
that the case of $\Lambda=m_{\star}$ 
looks unnatural from the theoretical point of view. 
The effective energy of the parton-level  collision ($\sqrt{\hat{s}}$) at the LHC 
is of the order of several TeV. In case $\sqrt{\hat{s}}>\Lambda$  the validity of series
expansion of the effective Lagrangian in terms of $\sqrt{\hat{s}}/\Lambda$
is arguable, at least for $m_{\star}$ values below 1~TeV.
With  respect to this fact, the choice of $\Lambda\simeq O(10$~TeV)
is safer and theoretically motivated. Here we will take $\Lambda=6$ TeV,
 corresponding to the ALEPH limit \cite{5}, as a reference. 

Although decay by contact interaction dominates for $\Lambda=m_{\star}$, the
decay via gauge interactions is proportional to $m_\star^3/\Lambda^2$, while
decay via contact interactions varies as $m_\star^5/\Lambda^4$. 
Therefore, the relative importance of the decay  mediated by contact 
interactions on the total decay width will be suppressed  by the 
factor $(m_\star/\Lambda)^2$.
This fact is clearly illustrated in Table~\ref{table1}. 
One can see that ( contrary to $\Lambda=m_{\star}$ )  in the case 
$\Lambda= 6$~TeV the excited electron decay is dominated by gauge interactions
up to $m_\star\leq 1.5$ TeV. However, for $m_\star\geq 3$ TeV the 
contribution from contact interactions to the total decay width is dominant 
and cannot be neglected. 
 Decays by
contact interactions yield $ejj$ and $ell$ final states, but the two jets or 
the two leptons accompanying the electron do not have the mass of the $Z$.

{\bf
\begin{center}
\begin{table}[ht]
\caption{Decay widths ( $\Gamma$ ) and branching ratios via gauge and contact
interactions, for $\Lambda=m_{\star}$ and $\Lambda=6$~TeV ( in parenthesis )
\label{table1}}
\begin{tabular}{cccc}
\hline $m_{\star }$(GeV) & $ \Gamma _{tot}($GeV$)$ & $\Gamma_{G}/\Gamma _{tot}$ &$ \Gamma _{C}/\Gamma_{tot}$\\ \hline
1000 &   89.9 (0.26) & 0.08 (0.75) & 0.92 (0.25) \\ \hline
3000 &    270 (20.8) & 0.08 (0.25) & 0.92 (0.75) \\ \hline
5000 &    451 (224)  & 0.08 (0.11) & 0.92 (0.89) \\ \hline
\end{tabular}
\end{table}
\end{center}
}

\section{Simulations and results}
Single production of excited electrons and their decays are
simulated. The  branching ratios in three
channels are given in Table~\ref{table2} as obtained from PYTHIA.
The simulations of excited lepton signal and relevant backgrounds were 
performed with COMPHEP\cite{16}, the COMPHEP-PYTHIA 
interface\cite{17} and PYTHIA\cite{18} programs chain.

As shown in the previous section,
the decay rate  of excited leptons via contact interactions is comparable to
the one via gauge interaction and can even  be the dominant one, contributing
up to 92\% to the total width ( $m_\star=6$~TeV case). 
Therefore it is very important to consider
gauge and contact decays of the excited electron simultaneously. In present
version of PYTHIA only gauge decay  of the excited fermions is
implemented. The use of PYTHIA $only$, therefore, for our task would be 
misleading.

\begin{table}[ht]
\caption{Branching ratios of excited lepton decay
via gauge and contact interactions.} 
\begin{tabular}{cccccc}
\hline $f=f^{\prime }$=1 &  & Gauge interactions \\
\hline Decay mode & e$^{\star }\rightarrow \nu $W $^{-}$ & e$^{\star
}\rightarrow$ e$\gamma $ & e$^{\star }\rightarrow eZ $ \\ \hline 
$ BR $ & $0.604$ & $0.283$ & $0.113$ \\ \hline
 &  & Contact interactions \\
\hline Decay mode & e$^{\star }\rightarrow$ e$\nu \nu$ & e$^{\star
}\rightarrow$ eee & e$^{\star }\rightarrow ejj $ \\ \hline
$ BR $ & $0.18$ & $0.11$ & $0.71$ \\ \hline
\end{tabular}
\label{table2}
\end{table}

One should also take into account the contact decays 
of the excited electron and, therefore, apply the proper correction for the 
gauge decay branching fraction (BF) from PYTHIA.
Taking this into account,
we have used PYTHIA with proper BF correction to perform decays of
excited electrons mediated by gauge interactions, while the
consideration of decays of excited leptons via contact interactions
were made available with the help of COMPHEP.

The ATLFAST\cite{19} code has been
used to take into account the experimental conditions prevailing at LHC for the 
ATLAS detector. The detector concept and its physics potential have been presented
in the Technical Proposal\cite{20} and the Technical Design Report\cite{21}.  The
ATLFAST program for fast detector simulations accounts for most of the detector
features: jet reconstruction in the calorimeters, momentum/energy smearing for
leptons and photons, magnetic field effects and missing transverse energy. It
provides a list of reconstructed  jets, isolated leptons and photons. In most
cases, the detector dependent parameters were tuned to values expected for the
performance of the ATLAS detector from full simulation.

The electromagnetic calorimeters were used to reconstruct the energy of leptons
in cells of dimensions $ \Delta \eta \X \Delta \phi =0.025\X 0.025 $
within the pseudorapidity range $ -2.5<\eta <2.5 $; $\phi$ is the azimuthal angle. 
The electromagnetic energy resolution is given by $ 0.1/\sqrt{E}(GeV)\bigoplus 0.007 $ 
over this pseudorapidity ($ \eta $) region.
The electromagnetic showers are identified as leptons when they lie within a cone
of radius $ \Delta R=\sqrt{(\Delta \eta)^{2} \X (\Delta \phi)^{2}} = 0.2 $ 
and possess a transverse energy $ E_{T}>5 $ GeV. 
Lepton isolation criteria were applied, requiring a distance $\Delta R>0.4$
from other clusters and maximum transverse energy deposition, $ E_{T}<10 $ GeV, 
in cells in a cone of radius $ \Delta R=0.2 $ around the direction of electron emission.

It must be mentioned that standard parametrization in the ATLFAST has been used
for the electron resolution but detailed studies are needed, using test beam 
data and GEANT full simulation to validate the extrapolation of the resolution function to electron energies in the TeV range.

\subsection{$e^{\star }\rightarrow e\gamma$ Channel}

The production cross sections for the excited electrons and backgrounds are given 
in Tables~\ref{table3} and ~\ref{table4}, respectively.

In order to enrich the event statistics in the region of high invariant masses, 
simulated background events were generated separately  
in different regions of the reaction transverse momentum, $\hat{p}_T$.
 
Several types of backgrounds were selected for comparison with the signal
events.

The natural background for the signal process studied is the process 
\qqb \ox $Z + \gamma$, where $Z$ decays to two leptons (electron - positron pair) 
yielding two leptons and a photon in the final state.

Another type of background process is the single production of $Z$ plus jet, where $Z$ decays
to the lepton pair and the additional jet in the event could be misidentified with a photon.
However, the photon/jet separation studies\cite{21} done in ATLAS give a jet total rejection of around 900 for jet $E_{T}$ around of 20 GeV at high luminosity ( $10^{34} cm^{-2} s^{-1}$ ). It is expected that the rejection
factor will be at least as high at higher energies.

\begin{center}
\begin{table}[h]
\caption{Cross section $\X$ BR (pb) (PYTHIA) for \qqb \ox $e^{*}e$ \ox $e^{+}e^{-}\gamma$, scale $\Lambda=m_{\star}$ and couplings $f=f^{\prime }=1$.}
\centering{
\begin{tabular}{ccccccc}
\hline
$m_{\star}$(GeV)&$ 500 $&$ 1000 $&$ 2000 $&$ 3000 $&$ 4000 $&$ 5000 $ \\ \hline
\qqb \ox $e^{*}e$ \ox $e^{+}e^{-}\gamma  $&$ 2950.$&$ 87.3 $&$ 1.35 $&$ 6.5\X 10^{-2} $&$ 4.25\X 10^{-3} $&$ 3.3\X 10^{-4}$\\ 
\hline
\end{tabular}}
\label{table3}
\end{table}
\end{center}

\begin{center}
\begin{table}[tbp]
\caption{Cross section $\sigma$(pb) from PYTHIA for the total background over various $\hat{p}_T$ (TeV) ranges}
\centering{
\begin{tabular}{ccccc}
\hline
$\hat{p}_T$           &$    \gamma+Z   $&$      Z+jet        $\\ \hline
$0.1<\hat{p}_T<0.3$   &$ 5.2\X 10^{-2} $&$    27.71          $\\
$0.3<\hat{p}_T<0.6$   &$ 1.3\X 10^{-3} $&$    5.1\X 10^{-1}  $\\ 
$0.6<\hat{p}_T<1.0$   &$ 7.4\X 10^{-5} $&$    2.0\X 10^{-2}  $\\
$\hat{p}_T>1.0$       &$ 5.5\X 10^{-6} $&$    6.4\X 10^{-3}  $\\ 
\hline
\end{tabular}}
\label{table4}
\end{table}
\end{center}

Decays of excited electrons produce characteristic event topologies and signatures
which allow one to distinguish them from their backgrounds. Due to their large
mass, excited electrons would produce events with large transverse energy. We study
the event topology of excited electrons in the decay channel 
$e^{\star}\rightarrow e\gamma$ and apply the selection criteria for both 
signal and background events. 

The following cuts were used to enhance the signal:

\begin{itemize}
\item The transverse momentum of leptons was required to be at least 100 GeV. 
\item Electrons were required to be within pseudorapidity acceptance of the ATLAS tracker and precise calorimetry, i.e. $|\eta |<2.5$.
\end{itemize}

\begin{figure}
\begin{center}
\includegraphics[width=12cm,height=12cm]{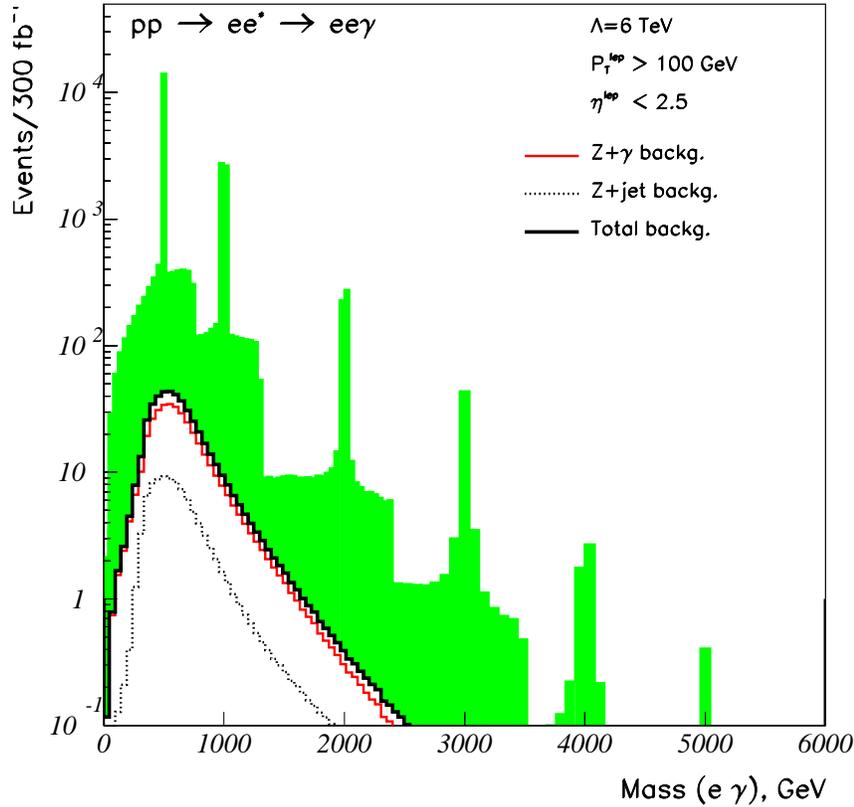}
\end{center}
\caption{Invariant mass distributions of $e\gamma$ for various excited lepton masses for an integrated 
luminosity of $300 fb^{-1}$}. 
\label{fig5}
\end{figure}

The resulting invariant mass distributions of photon - electron pairs are presented in
Fig.~\ref{fig5} for different $ m_{\star}$ masses of the excited lepton for $\Lambda=6$ TeV. 
The resonances are clearly seen above the combinatorial and other backgrounds.
The distributions were normalized to an integrated luminosity of $L=300$ fb$^{-1}$.

The signal significances are defined at each mass point as $S/\sqrt{B}$ 
where $ S $ and $ B $ are the number of accepted signal and
background events in the selected mass bin width ( $\Delta M$ ), respectively.
For the mass bin width the value was taken equal to $\pm 2\sigma$ width of the 
invariant mass distribution around the excited electron peak. 
A modest rejection factor of 1000\cite{21} has been used for
jets mistagged as photons to suppress  the $Z + jet$ background.
Achievable mass limits for the excited single electron production are 
established by requiring that at least 5 events survive the cuts and $S/\sqrt{B}>5$. 

As can be seen from Table ~\ref{table5}, 
the masses achieved for excited single lepton in the photon and electron 
decay modes are $\sim 4.0 $ TeV for $\Lambda=6$ TeV, the couplings 
$f=f'=1$ and for an integrated luminosity $L=300$ fb$^{-1}$ at CERN LHC.

\begin{center}
\begin{table}[tbp]
\caption{Signal significance $S/\sqrt{B}$, 
$S$ for signal, $B$ for total background, and number of events are 
calculated in the window $\Delta M$ around the excited electron peak 
for an integrated luminosity of $L=300 fb^{-1}$ and $\Lambda=6$ TeV.}
\centering{
\begin{tabular}{lllllll} \hline
$m_{\star}$(TeV) \ox &   0.5   &  1.0  &  2.0  &  3.0  &  4.0 &\\ \hline
$\Delta M$, GeV      &   16    &   30  &   50  &   80  &  120 &\\ \hline
$S$                  &  13700  &  5281 &  490  &   43  &  5.0 &\\ \hline
$S/\sqrt{B}$         &  2385   &  2104 & 1044  &  311  &  66  &\\ \hline
\end{tabular}}
\label{table5}
\end{table}
\end{center}

\subsection{$e^{\star }\rightarrow Ze$  / eejj(eeee) channels: combining gauge and contact decays}

Here we consider the decay of excited electrons to $Ze$ mediated by gauge 
interactions together with the
3-body decay ($ell$, $ejj$) of the excited electron mediated by contact 
interactions.
It is natural to treat those decays together since we consider 
the vector boson decay modes to $Z \rightarrow ee$ or $Z \rightarrow qq$,
and  the signal consists of four electrons or two jets and two electrons.
The decay channel $Z$ \ox $\mu\mu$ is expected to give results comparable to $Z$ \ox $ee$, 
but with somewhat worse energy resolution.

We will allow for the possibility that one of the electrons could 
be outside the acceptance region of the detector or could be lost by some inefficiency of reconstruction.

The dominant backgrounds for the signal process studied when $Z$ decays to an
electron-positron pair are:

\begin{itemize}

\item \qqb \ox $Z + \gamma$, where $Z$ decays to two leptons (electron - positron pair)
yielding two leptons and a photon in the final state. Good rejection of photons
in electron reconstruction could help to suppress this source of background.

\item $Z+Z$ production, where both $Z$ decay to an electron
and a positron.

\item $W+Z$ production, where $Z$ again decays to an electron
and a positron, and $W$ decays to an electron and a neutrino.

\end{itemize}

In the case when $Z$ decays to jets the following backgrounds could be relevant:

\begin{itemize}

\item $Z+jet$ production, where a $Z$ decay to an electron
and a positron, and additional jet produced by means of the final state radiation mechanism.

\item $Z+Z$ production, where one of $Z$ decays to an electron-positron pair, while the other $Z$ decays to jets.

\item \qqb \ox $Z + \gamma$, where $Z$ decays to two jets
yielding two jets and a photon in the final state. 
Again, as in the case of $Z \rightarrow ee$ decay, good rejection of photons in electron reconstruction could help to suppress this source of background.

\end{itemize}

The production cross sections for the signal channels and
backgrounds are given in Table ~\ref{table6} and Table
~\ref{table7}, respectively.

\begin{center}
\begin{table}[ht]
\caption{Cross section $\X$ BR (pb) from PYTHIA for $e^{*}e$ \ox
$Zee$, scale $\Lambda=m_{\star}$ and couplings $f=f^{\prime }=1$.
} \centering{
\begin{tabular}{cccccc}
\hline $m_{\star}(GeV)$&$ 500 $&$ 1000 $&$ 2000 $&$ 3000 $&$ 4000 $\\ \hline
Z \ox ee &$  38.5 $&$  1.2 $&$ 1.9\X 10^{-2} $&$ 8.5\X 10^{-4} $&$ 5.5\X 10^{-5} $\\
Z \ox jj &$ 800.  $&$ 24.4 $&$ 3.8\X 10^{-1} $&$ 1.8\X 10^{-2} $&$ 1.2\X 10^{-3} $\\
\hline
\end{tabular}} \label{table6}
\end{table}


\begin{table}[ht]
\caption{Cross section $\sigma$(pb) from PYTHIA for the total background
with $Z$ \ox $ee$ ( $jj$ ) over given $\hat{p}_T$ (TeV) ranges}
\centering{
\begin{tabular}{ccccc}
\hline
$Z$ \ox $ee$ & & & \\ \hline
$\hat{p}_T$           &    $Z+\gamma$   &       $Z+Z$     &     $Z+W$       \\ 
\hline
$0.1<\hat{p}_T<0.3$   &$ 5.2\X 10^{-2} $&$ 1.7\X 10^{-3} $&$ 1.1\X 10^{-2}$ \\
$\hat{p}_T>0.3$       &$ 1.4\X 10^{-3} $&$ 5.8\X 10^{-5} $&$ 3.0\X 10^{-4}$ \\
\hline \hline
$Z$ \ox $jj$ & & & \\ \hline
$\hat{p}_T$          &    $Z + jet$    &   $Z+\gamma$   &       $Z+Z$     \\
\hline
$0.1<\hat{p}_T<0.3$  &$   21.55       $&$ 9.7\X 10^{-1} $&$ 7.4\X 10^{-1}$ \\
$\hat{p}_T>0.3$      &$ 4.0\X 10^{-1} $&$ 2.7\X 10^{-2} $&$ 2.6\X 10^{-2}$ \\
\hline
\end{tabular}}
\label{table7}
\end{table}
\end{center}

The following cuts were used to separate the signal from background:

\begin{itemize}

\item $eeee$ final state case
 
\begin{itemize}
\item The transverse momenta of leptons were required to be at least 
60 GeV.
\item Electrons were required to be within pseudorapidity acceptance of the ATLAS 
tracker and precise calorimetry, i.e. $|\eta |<2.5$.
\end{itemize}

\item $eejj$ final state case

\begin{itemize}
\item The transverse momenta of jets and leptons were required to be at least 
120 GeV.
\item Electrons were required to be within pseudorapidity acceptance of the ATLAS 
tracker and precise calorimetry, i.e. $|\eta |<2.5$.
\item The cosine of the opening angle between an electron and a jet was required to be greater than -0.8. 
\end{itemize}

\end{itemize}

\begin{figure}[ht]
\begin{center}
\begin{tabular}{c c c}
\hspace{-2.0cm}
\includegraphics[width=12cm,height=12cm]{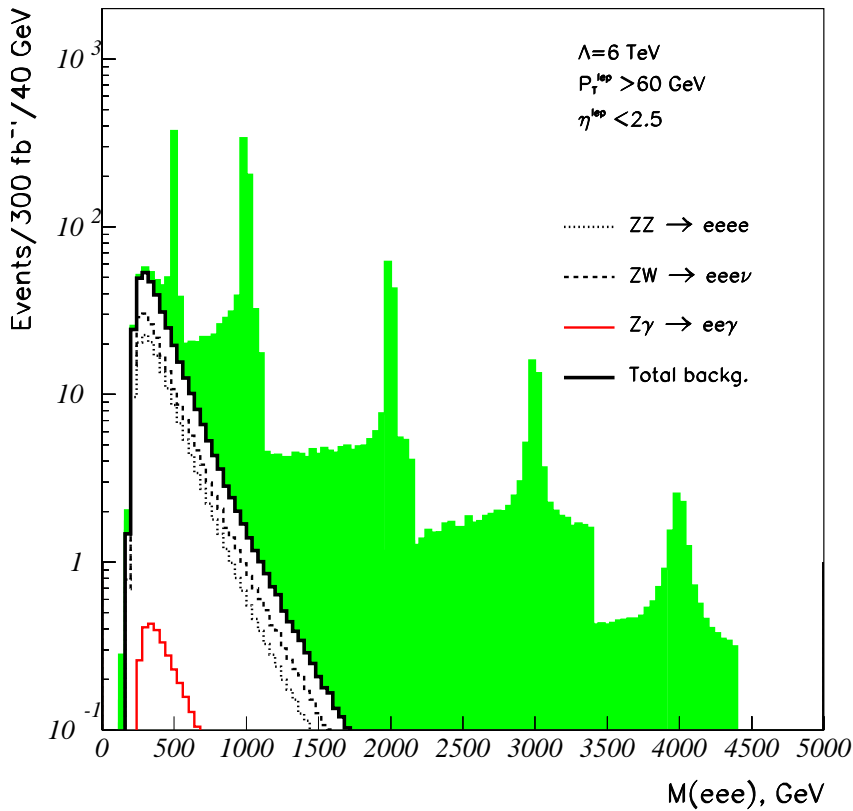}
\end{tabular}
\end{center}
\caption{\label{fig6}
Invariant mass distributions of combined  $Ze$ ($Z$ \ox ee) + $eeee$
contributions from gauge and contact excited lepton decays for
masses of 500 GeV, 1 TeV, 2 TeV, 3 TeV and 4 TeV for an integrated
luminosity of $300 fb^{-1}$ and $\Lambda = 6$ TeV.}
\end{figure}

The resulting invariant mass distributions of three lepton system
are presented in Fig.~\ref{fig6} for different $ m_{\star} $ masses 
of the excited lepton. The four electrons with highest \pT were selected for
the reconstruction. As before, a scale $\Lambda = 6$ TeV is taken as 
reference. 
A photon/electron rejection factor of 500\cite{21} has been used to
suppress  the $Z + \gamma$ background.
The distributions were normalized to an integrated luminosity of $L=300$ fb$^{-1}$.

For a $Z$ boson that decays hadronically, the resulting jets are highly boosted
for large values of $m_{\star}$. If the two jets were merged as a single jet, 
the mass of the excited lepton was reconstructed as the invariant mass of the 
jet-electron system.

\begin{figure}[ht]
\begin{center}
\begin{tabular}{ccc}
\hspace{-2.0cm}
\includegraphics[width=12cm,height=12cm]{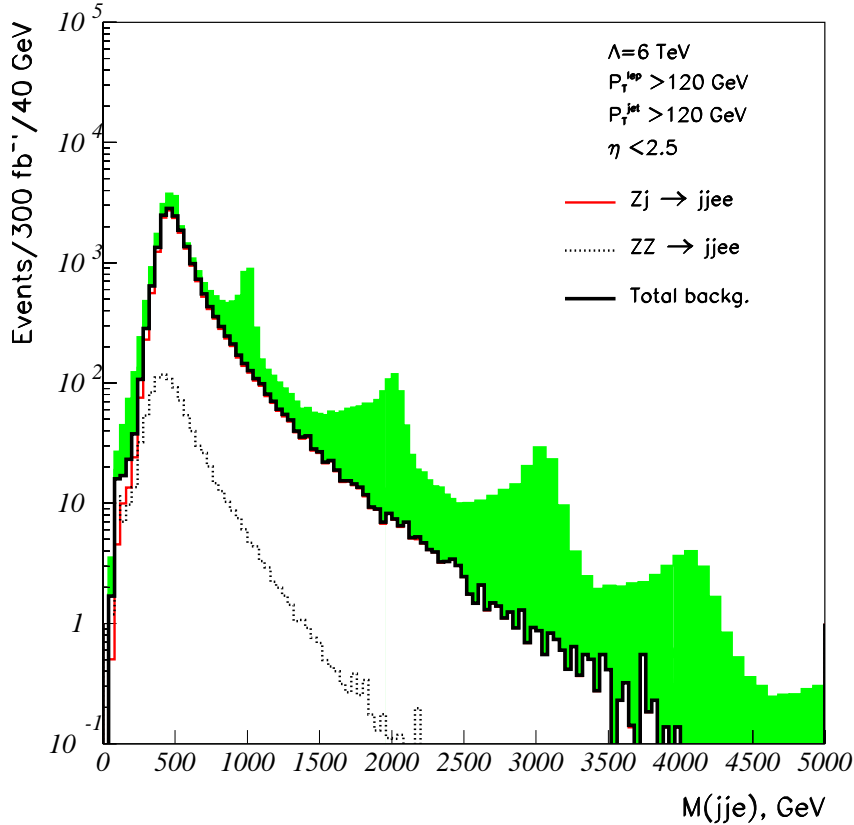}
\end{tabular}
\end{center}
\caption{Invariant mass distributions of $Ze$ ( $Z$ \ox jj ) + $eejj$ of combined contributions from gauge and contact excited lepton decays
for masses of 500 GeV, 1 TeV, 2 TeV, 3 TeV and 4 TeV for an integrated luminosity of $300 fb^{-1}$ and $\Lambda = 6$ TeV.
\label{fig7}}
\end{figure}

The resulting invariant mass distributions of a lepton and a jet
pair are presented in Fig.~\ref{fig7} for different $ m_{\star} $
 and $\Lambda = 6$ TeV. The same photon/electron rejection factor of 500 has been used to suppress the 
$Z + \gamma$ background. 
The distributions were normalized to an integrated luminosity of $L=300$ fb$^{-1}$.
 For $m_{\star}=500$~GeV we have the peak  falling  on top of
 background. However, 
 this is not the problem, since signal to background ratio is about one.
 So, for the signal which is  about the same as the background the problem
 of  understanding of K-factor does not occur.
 We believe that  by the time of LHC, uncertainties for the
$Wjj$ cross section will be known far better then 50\%.

The signal significances are defined at each mass point as $S/\sqrt{B}$
where $ S $ and $ B $ are the number of accepted signal and
background events in the selected mass bin, respectively.
Achievable mass limits for the excited single electron production are
established by requiring that at least 5 events survive the cuts and
$ S/\sqrt{B}>5 $.

\begin{figure}[ht]
\begin{center}
\begin{tabular}{cc}
\hspace{-2.0cm}
\includegraphics[width=7.5cm,height=7.5cm]{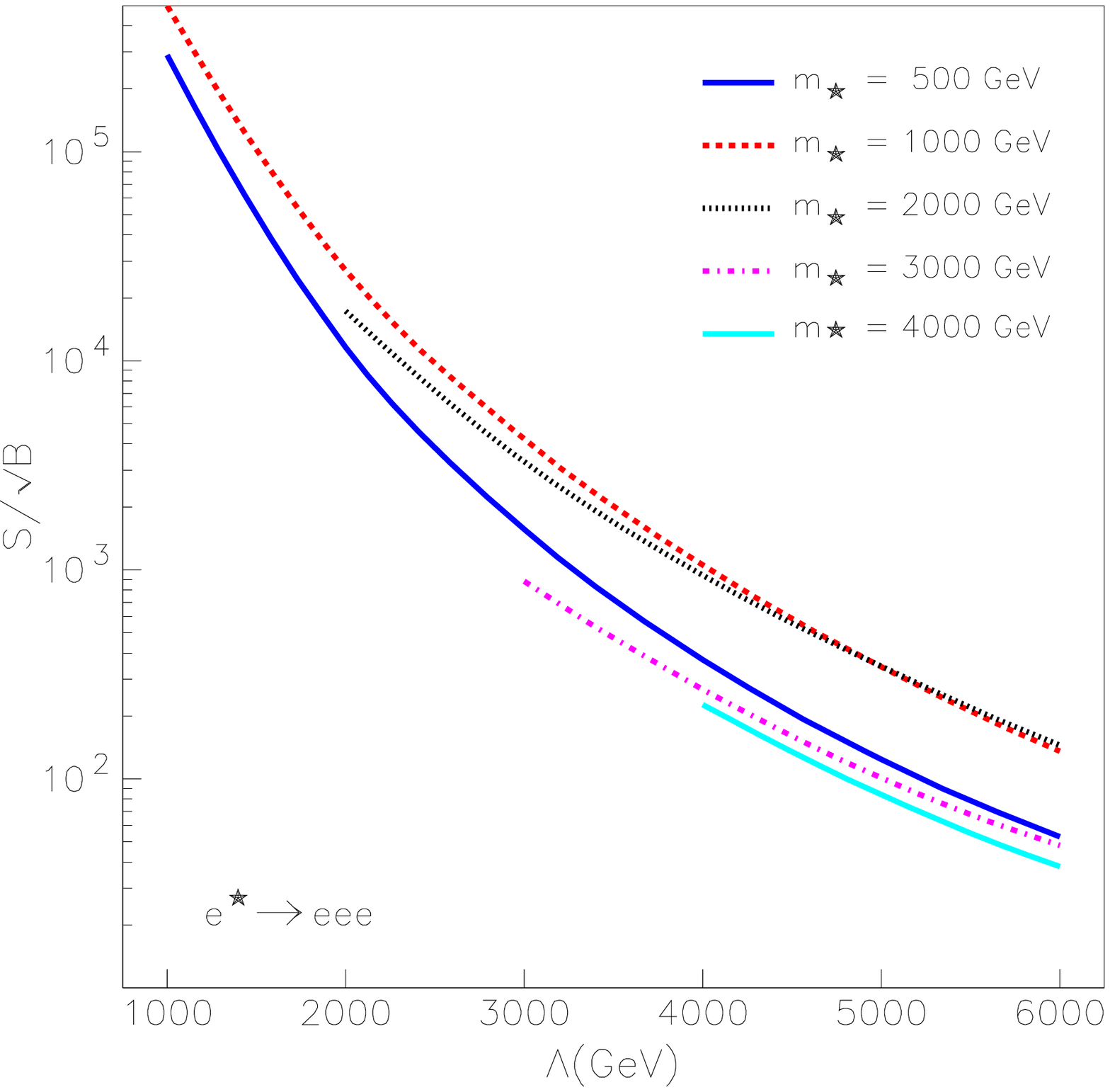}
\includegraphics[width=7.5cm,height=7.5cm]{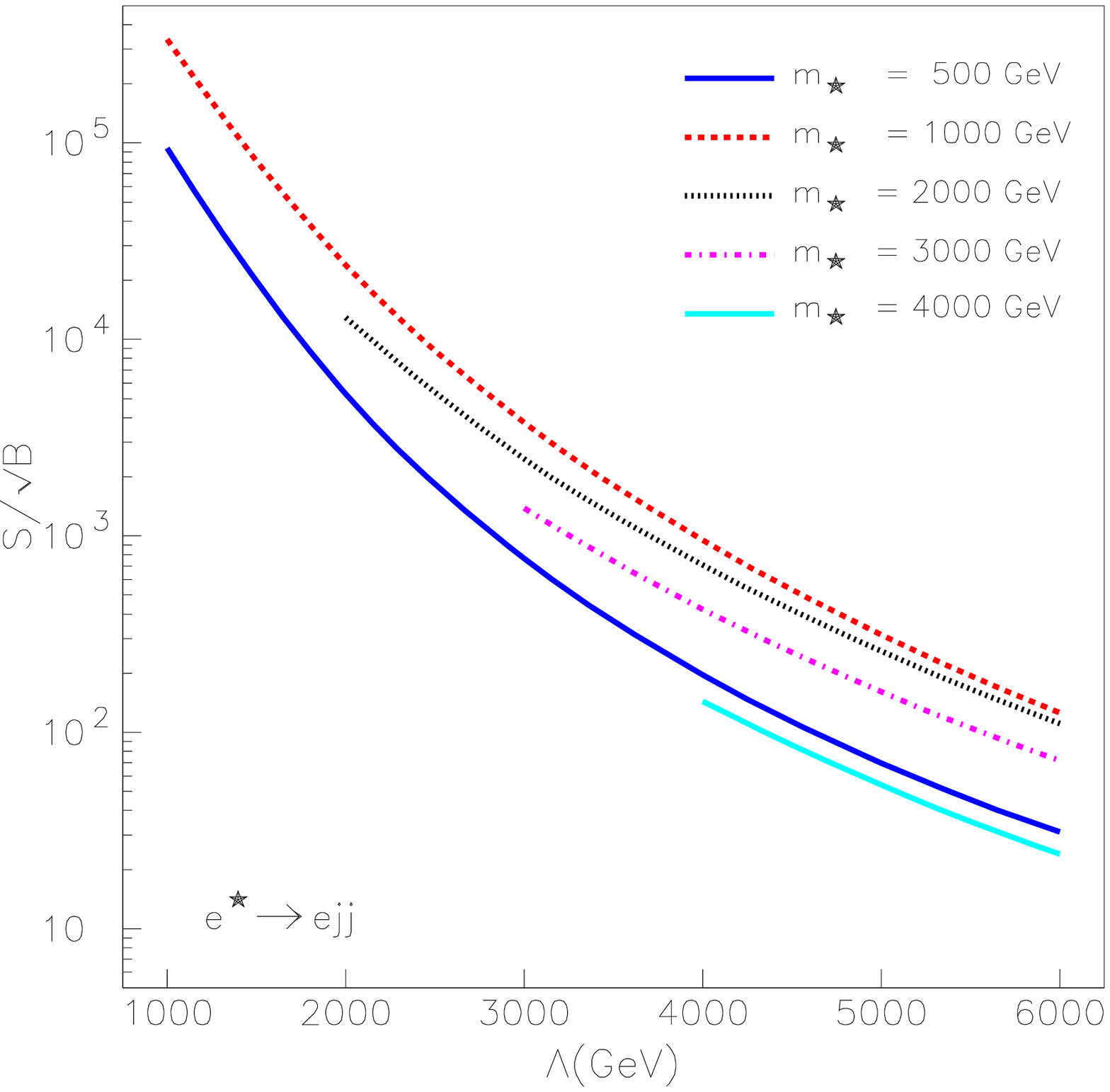}
\end{tabular}
\caption{Excited electron signal significance shown for different scale
$\Lambda$ at an integrated luminosity at LHC $L=300 fb^{-1}$. 
\label{fig8}}
\end{center}
\end{figure}

As can be seen from Table~\ref{table8} and Fig.~\ref{fig8},
the masses achieved for excited single lepton in the $Ze$
decay modes are $\sim 3.0 $ TeV for $\Lambda=6$ TeV, the couplings
$f=f'=1$ and for an integrated luminosity $L=300$ fb$^{-1}$ at CERN LHC.
Due to a low number of events in the signal events sample, we do not present
the data for the masses beyond 4 TeV, but with an extended data taking period,
this region of masses may also be reachable.

\begin{center}
\begin{table}[ht]
\caption{Signal significance $S/\sqrt{B}$,
$S$ for signal, $B$ for total background, and number of events are
calculated for an integrated luminosity of $L=300 fb^{-1}$ and
$\Lambda = 6$ TeV within selected mass bin width ($\Delta M$).}
\centering{
\begin{tabular}{llllll} \hline
$m_{\star}$(TeV)$\rightarrow$ &  0.5  &  1.0   &  2.0  &  3.0  &  4.0  \\ \hline
$q\overline{q}$ \ox $e^{\star }e$ \ox $eeee$ & & & &                   \\ 
\hline $\Delta M$, GeV        &   20  &   38   &   63   &   84 &  120  \\ 
\hline $S$                    &  168  &  192   &   46   &   14 &  3.4  \\
\hline $S/\sqrt{B}$           &   53  &  135   &  145   &   99 &  76  \\ 
\hline \hline 
$q\overline{q}$ \ox $e^{\star }e$ \ox $eejj$  & & & &                  \\ 
\hline $\Delta M$, GeV        &   40  &   60   &  106  &  180  & 200   \\
\hline $S$                    & 2102  &  2415  &  636  &  176  &  24   \\
\hline $S/\sqrt{B}$           &   30  &   120  &  105  &   69  &  30   \\
\hline
\end{tabular}}
\label{table8}
\end{table}
\end{center}

\subsection{$e^{\star }\rightarrow W\nu$ Channel}
\begin{figure}[ht]
\begin{center}
\begin{tabular}{c c}
\hspace{-2.0cm}
\includegraphics[width=10cm,height=10cm]{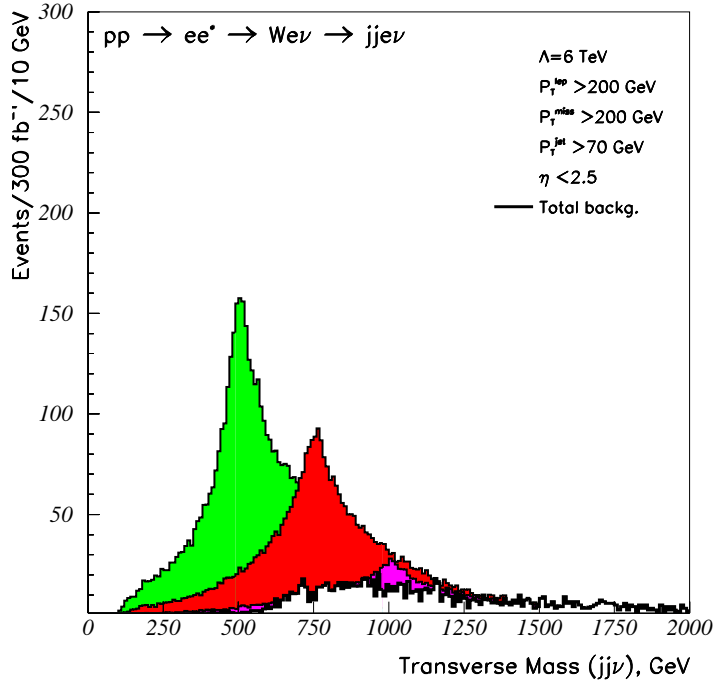}
\end{tabular}
\caption{Invariant mass distribution of ($jj\nu$) system 
for excited lepton masses of 500, 750, 1000 and 1250 GeV
for an integrated luminosity of $300 fb^{-1}$ and $\Lambda = 6$ TeV. 
The total background from $W+jet$ and $W+W$  production is also shown.
\label{fig9}}
\end{center}
\end{figure}

The production cross sections for this channel are
given in Table~\ref{table9}.
\begin{center}
\begin{table}[ht]
\caption{Cross section $\X$ BR (pb) (PYTHIA) for $e^{*}e$ \ox
$W+\nu e$, scale $\Lambda=m_{\star }$ and couplings $f=f^{\prime
}=1$. } \centering{
\begin{tabular}{cccccc}
\hline
$m_{\star }(GeV)$&$ 1000. $&$ 2000.$&$ 3000. $&$ 4000.$&$ 5000. $ \\ \hline
$W$ \ox $jj$&  $ 93.5 $&$ 1.1 $&$ 3.9\X 10^{-2} $&$ 2.0\X 10^{-3} $&$ 1.3\X 10^{-4}$\\
$W$ \ox $e\nu$&$ 14.8 $&$ 0.2 $&$ 6.0\X 10^{-3} $&$ 3.3\X 10^{-4} $&$ 2.0\X 10^{-5}$\\
\hline
\end{tabular}}
\label{table9}
\end{table}
\end{center}
In the subprocesses involving a $W$ decaying hadronically,
the final state consists of two jets, an electron and a neutrino. 
An attempt to reconstruct the mass of ($jj\nu$) system is presented in
Fig.~\ref{fig9} for excited electron masses of 500, 750, 1000, 1250 and 
1500 GeV for an integrated luminosity of $300 fb^{-1}$ and $\Lambda = 6$ TeV.
The two jets with highest \pT were used for the mass reconstruction.
Because the reconstructed mass cannot account for the longitudinal missing
momentum, a wide distribution is obtained.

The plots were obtained using following cuts:

\begin{itemize}
\item The transverse momenta of jets were required to be at least 70 GeV.
\item The transverse momentum of electron was required to be 200 GeV 
within pseudorapidity $|\eta |<2.5$ for all excited lepton masses considered.
\item It was required to have a reconstructed $W$ \ox $jj$ with a mass in the 
$70-90$ GeV range in the event to be  selected.
\item The missing \pT cut was required to be 200 GeV.
\end{itemize}

As can be seen from Fig. ~\ref{fig9}, signal signatures can be
observed for excited electron masses from 500 GeV  up to 1000 GeV
over the total background.

The second subprocess, when $W$ decays to a lepton
and a neutrino, gives pairs of leptons and neutrinos. The
effective mass reconstruction is very difficult
in this case. It requires the evaluation of the neutrino longitudinal
momenta, which is impossible without certain kinematical
constraints. 
 However, estimating the acceptance and 
 efficiencies at 65\% and 20\%, respectively,  the transverse mass distribution
 $ m_T = \sqrt{|\vec{p}_T(e)|^2 + |\vec{p}_T(\ell) + \vec{p}_T^{miss}|^2 }$
($e$ is the electron recoiling against $e^*$ and $\ell$ is the  lepton
 from $W$ decay)  should yield an observable excess of events over
 backgrounds for excited electron masses up to 4.5 TeV. This process could
 therefore confirm  excited electron observation, obtained from others channels.

\section{Conclusions}
In the framework of a composite model of quarks and leptons, sharing common constituents,
excited electrons could be produced copiously at CERN LHC. The first
indication for the excited electrons could be the production of ordinary leptons at rates much larger than expected in the framework of the SM.  
Large lepton yields are expected if quarks and
leptons share common sub-constituents. Clean signatures are predicted by the large fraction
of decays with leptons in the final state, that are a consequence of the uniform coupling
among quarks and leptons through contact interactions. We have presented the results of
excited single electron production with  subsequent decay mediated  by gauge and contact
interactions. Singly produced excited electrons could be accessible up to a mass of 5 TeV  at LHC for $\Lambda = 6$ TeV. For the excited electrons masses decay
both contact and gauge interactions plays important role and 
should be considered together.
For comparatively light excited leptons ($<1$~TeV) the
decays mediated by contact interactions may be considerably suppressed, while for higher
masses of excited states the contribution from contact interactions to the total decay
width is dominant. The decay channel $Z$ \ox $\mu\mu$ is expected
to give results comparable to $Z$ \ox $ee$, but with somewhat lower energy resolution.

\section{Acknowledgments}
This work has been performed within the ATLAS Collaboration with the help of the simulation
framework and tools which are the result of the collaboration-wide efforts.
We would like also to thank G. Azuelos, H.Baer, D. Froidevaux, F. Gianotti, I. Hinchliffe, L. Poggioli, L.Reina for their comments about the subject. 
C.L. and R.M. thank NSERC/Canada for their support.
\end{document}